\documentclass[11pt]{article}
\usepackage{amssymb,amsfonts}
\usepackage{graphics, amsmath}
\usepackage{subfigure}
\usepackage{graphicx}
\usepackage{verbatim}
\usepackage{color}
\usepackage{hyperref}
\hypersetup{colorlinks}

\definecolor{darkred}{rgb}{1,0,0}
\definecolor{darkgreen}{rgb}{0,0.5,0}
\definecolor{darkblue}{rgb}{0,0,1}
\definecolor{orange}{rgb}{1,0.5,0}
\definecolor{green}{rgb}{0,1,0}
\definecolor{purple}{rgb}{.5,0,1}

\hypersetup{ colorlinks,
linkcolor=darkred,
filecolor=darkgreen,
urlcolor=darkblue,
citecolor=darkblue,
linktocpage=true }

\definecolor{markcolor}{rgb}{.25,0,1}

\definecolor{markcolor2}{rgb}{1,0,0}

\definecolor{markcolor3}{rgb}{0,1,0}

\def\hybrid{\topmargin 10pt    \oddsidemargin 0.20in 
        \headheight 0pt \headsep 0pt
        \textwidth 15.2cm      
        \textheight 21.0cm       
        \marginparwidth 1.075in
        \parskip 5pt plus 1pt   \jot = 1.5ex}

\hybrid

\catcode`\@=11

\def\marginnote#1{}
\newcount\hour
\newcount\minute
\newtoks\amorpm
\hour=\time\divide\hour by60 \minute=\time{\multiply\hour by60
\global\advance\minute by-\hour}
\edef\standardtime{{\ifnum\hour<12 \global\amorpm={am}%
        \else\global\amorpm={pm}\advance\hour by-12 \fi
        \ifnum\hour=0 \hour=12 \fi
        \number\hour:\ifnum\minute<10 0\fi\number\minute\the\amorpm}}
\edef\militarytime{\number\hour:\ifnum\minute<10 0\fi\number\minute}

\def\draftlabel#1{{\@bsphack\if@filesw {\let\thepage\relax
   \xdef\@gtempa{\write\@auxout{\string
      \newlabel{#1}{{\@currentlabel}{\thepage}}}}}\@gtempa
   \if@nobreak \ifvmode\nobreak\fi\fi\fi\@esphack}
        \gdef\@eqnlabel{#1}}
\def\@eqnlabel{}
\def\@vacuum{}
\def\draftmarginnote#1{\marginpar{\raggedright\scriptsize\tt#1}}

\def\draft{\oddsidemargin -.5truein
        \def\@oddfoot{\sl preliminary draft \hfil
        \rm\thepage\hfil\sl\today\quad\militarytime}
        \let\@evenfoot\@oddfoot \overfullrule 3pt
        \let\label=\draftlabel
        \let\marginnote=\draftmarginnote
   \def\@eqnnum{(\theequation)\rlap{\kern\marginparsep\tt\@eqnlabel}%
\global\let\@eqnlabel\@vacuum}  }

\def\draft2{
        \def\@oddfoot{\sl preliminary draft \hfil
        \rm\thepage\hfil\sl\today\quad\militarytime}
        \let\@evenfoot\@oddfoot \overfullrule 3pt
        \let\label=\draftlabel
        \let\marginnote=\draftmarginnote
   \def\@eqnnum{(\theequation)\rlap{\kern\marginparsep\tt\@eqnlabel}%
\global\let\@eqnlabel\@vacuum}  }

\def\preprint{\twocolumn\sloppy\flushbottom\parindent 2em
        \leftmargini 2em\leftmarginv .5em\leftmarginvi .5em
        \oddsidemargin -.5in    \evensidemargin -.5in
        \columnsep .4in \footheight 0pt
        \textwidth 10.in        \topmargin  -.4in
        \headheight 12pt \topskip .4in
        \textheight 6.9in \footskip 0pt
        \def\@oddhead{\thepage\hfil\addtocounter{page}{1}\thepage}
        \let\@evenhead\@oddhead \def\@oddfoot{} \def\@evenfoot{} }

\def\numberbysection{\@addtoreset{equation}{section}
        \def\theequation{\thesection.\arabic{equation}}}

\def\underline#1{\relax\ifmmode\@@underline#1\else
        $\@@underline{\hbox{#1}}$\relax\fi}

\def\titlepage{\@restonecolfalse\if@twocolumn\@restonecoltrue\onecolumn
     \else \newpage \fi \thispagestyle{empty}\c@page\z@
        \def\thefootnote{\fnsymbol{footnote}} }

\def\endtitlepage{\if@restonecol\twocolumn \else \newpage \fi
        \def\thefootnote{\arabic{footnote}}
        \setcounter{footnote}{0}}  

\catcode`@=12 \relax

\def\figcap{\section*{Figure Captions\markboth
        {FIGURECAPTIONS}{FIGURECAPTIONS}}\list
        {Figure \arabic{enumi}:\hfill}{\settowidth\labelwidth{Figure
999:}
        \leftmargin\labelwidth
        \advance\leftmargin\labelsep\usecounter{enumi}}}
 \relax
\def\tablecap{\section*{Table Captions\markboth
        {TABLECAPTIONS}{TABLECAPTIONS}}\list
        {Table \arabic{enumi}:\hfill}{\settowidth\labelwidth{Table
999:}
        \leftmargin\labelwidth
        \advance\leftmargin\labelsep\usecounter{enumi}}}
 \relax
\def\reflist{\section*{References\markboth
        {REFLIST}{REFLIST}}\list
        {[\arabic{enumi}]\hfill}{\settowidth\labelwidth{[999]}
        \leftmargin\labelwidth
        \advance\leftmargin\labelsep\usecounter{enumi}}}
 \relax
\makeatletter
\newcounter{pubctr}
\def\publist{\@ifnextchar[{\@publist}{\@@publist}}
\def\@publist[#1]{\list
        {[\arabic{pubctr}]\hfill}{\settowidth\labelwidth{[999]}
        \leftmargin\labelwidth
        \advance\leftmargin\labelsep
        \@nmbrlisttrue\def\@listctr{pubctr}
        \setcounter{pubctr}{#1}\addtocounter{pubctr}{-1}}}
\def\@@publist{\list
        {[\arabic{pubctr}]\hfill}{\settowidth\labelwidth{[999]}
        \leftmargin\labelwidth
        \advance\leftmargin\labelsep
        \@nmbrlisttrue\def\@listctr{pubctr}}}
 \relax
\makeatother

\def\be{\begin{equation}}
\def\ee{\end{equation}}
\def\ba{\begin{eqnarray}}
\def\ea{\end{eqnarray}}

\newcommand{\ph}[1]{\phantom{#1}}

\def\no{\noindent}

\def\IR{\relax{\rm I\kern-.18em R}}

\newcommand{\e}[1]{\ensuremath{\text{e}^{#1}}}

\def\aa{a}
\def\ab{\bar{a}}
\def\aav{b}
\def\abv{\mathchar'26\mkern-9mu b}
\def\zz{z}
\def\zb{\bar{z}}
\def\zzX{y}
\def\zbX{\bar{y}}
\def\dd{\mathrm{d}}
\newcommand{\Pexp}[1]{\ensuremath{\mathrm{P\, exp} \left\lbrace #1 \right\rbrace}}

\def\bse{\begin{small}\begin{equation*}}
\def\ese{\end{equation*}\end{small}}

\begin{document}


\renewcommand{\theequation}{\thesection.\arabic{equation}}
\csname @addtoreset\endcsname{equation}{section}

\newcommand{\eqn}[1]{(\ref{#1})}

\begin{titlepage}
\begin{center}
\strut
\vskip 1.5cm

{\Large \bf Space $\&$ time discontinuities in Liouville theory and the deformed oscillator model}

\vskip 0.5in

{{\bf Anastasia Doikou and Iain Findlay}} \vskip 0.2in

 \vskip 0.02in
{\footnotesize
 Department of Mathematics, Heriot-Watt University,\\
EH14 4AS, Edinburgh, United Kingdom}
\\[2mm]

\vskip .1cm


{\footnotesize {\tt E-mail: a.doikou@hw.ac.uk, iaf1@hw.ac.uk}}\\

\end{center}

\vskip 1.0in

\centerline{\bf Abstract}

We consider the deformed harmonic oscillator as a discrete version of the Liouville theory and study this model in the presence of local integrable defects. From this, the time evolution of the defect degrees of freedom are determined, found in the form of the local equations of motion. We also revisit the continuous Liouville theory, deriving its local integrals of motion and comparing these with previous results from the sine-Gordon point of view. Then, the generic B\"{a}cklund type relations are presented, corresponding to the implementation of time-like and space-like impurities in the continuum model. Finally, we consider the interface of the Liouville theory with the free massless theory. With the appropriate choice of the defect (Darboux) matrix we are able to derive the hetero-B\"acklund transformation for the Liouville theory.
\no

\vfill

\end{titlepage}
\vfill \eject


\tableofcontents

\-

\section{Introduction}

There has been a considerable amount of work in recent years devoted to the study of local defects in discrete and continuum integrable models \cite{delmusi}--\cite{doikou-backl}. Here, we continue this line of investigation considering the Liouville theory and its discrete integrable analogue, mainly focusing on the equations of motion, equivalent to the time evolution of the defect degrees of freedom, which are similar to B\"{a}cklund type relations.

We choose to consider this type of models mainly due to the singular nature of the associated Lax matrices. We follow the Hamiltonian description, but also utilise the idea that the integrable defect may be seen as a quasi-B\"{a}cklund transformation \cite{doikou-backl}. We consider here a Lax operator that is a special limit of the (discrete) sine-Gordon Lax operator (see \cite{limit1, limit2}), and it is modified in a similar fashion to \cite{corrigan-atft, aguire} such that it may be easily employed at the quantum level, and in particular, within the Bethe ansatz formulation. Such models are associated to harmonic oscillator algebras or suitable deformations thereof. Other related models are, for instance, the Ablowitz-Ladik and the non-linear Schr\"{o}dinger models \cite{ablow}. The ultimate aim is to deal with all these models under a unified algebraic frame as far as the B\"{a}cklund transformation relations are concerned.

More specifically, we shall consider here both the discrete and continuous versions of the Liouville model. The Lax operator $ L $ of the discrete model has the form:
\begin{equation}
	L_n(\lambda) = \left( \begin{matrix}
		\e{\lambda} v_n - \e{-\lambda} v_n^{-1} & \ab_n \\
		\aa_n & -\e{-\lambda} v_n
	\end{matrix} \right), \label{eq:lax1}
\end{equation}

\noindent at each site $ n $, where $ \aa_n $, $ \ab_n $, and $ v_n $ are the fields, and $ \lambda $ is the spectral parameter.

We shall briefly review below how a suitably subtle continuum limit of \eqref{eq:lax1} provides the Lax operator of the continuous Liouville theory (see also \cite{limit1}). Specifically, we multiply the matrix above with the anti-diagonal $ 2\times 2 $ matrix $ \sigma^x $ and consider the following map:
\begin{equation}
	a_n = \e{-\frac{i\pi_n}{2}}, ~~~~~\bar a_n = \e{\frac{i\pi_n}{2}}, ~~~~v_n = \e{-\frac{i\phi_n}{2}}.
\end{equation}

Taking a suitable continuum limit:
\begin{equation}
	\phi_n \to \phi(x), ~~~~\pi_n \to \pi(x),
\end{equation}

\noindent one obtains the Lax operator of the Liouville model:
\begin{equation}
	U(\lambda) = \frac{1}{2} \left( \begin{matrix}
		-i\pi & -2\e{-\lambda - i\phi} \\
		4\, \text{sinh}(\lambda - i\phi) & i\pi
	\end{matrix} \right),
\end{equation}
where $\phi,\ \pi$ are canonical classical fields.

The relevant $ r $-matrix that satisfies the classical Yang-Baxter equation is:
\begin{equation}
	r_{ab}(\lambda) = \frac{1}{\sinh{\lambda}} \left( \begin{matrix}
		\cosh{\lambda} & 0 & 0 & 0 \\
		0 & 0 & 1 & 0 \\
		0 & 1 & 0 & 0 \\
		0 & 0 & 0 & \cosh{\lambda}
	\end{matrix} \right),
\end{equation}

\noindent and we require that $ L_n $ satisfies a quadratic algebraic relation:
\begin{equation}
	\Big\lbrace L_{an}(\lambda),\ L_{bm}(\mu) \Big\rbrace = \big[ r_{ab}(\lambda - \mu),\ L_{an}(\lambda) L_{bm}(\mu) \big] \delta_{nm}. \label{eq:fund1}
\end{equation}

Similarly for the continuous Liouville model the $ U $ operator satisfies the following linear algebra:
\begin{equation}
	\Big \{U_a(\lambda, x),\ U_b(\mu, y) \Big \} = \Big [r_{ab}(\lambda -\mu),\ U_a(\lambda, x)+ U_b(\mu, y)\Big ] \delta(x - y). \label{eq:fund2}
\end{equation}

We shall first review the one dimensional discrete classical model, before implementing local integrable defects. After extracting the local integrals of motion we shall also find the time component of the Lax pair. Having these expressions at our disposal, we come to the main focus of the present analysis: the derivation of the time evolution of the degrees of freedom of the defect encoded on a suitable local Lax operator. In the case of the continuous Liouville model in the presence of defects we discuss the integrals of motion as well as the construction of the Lax pair, and verify their validity by comparing the massless limit of previous results in this context of the sine-Gordon model. We then focus on the B\"{a}cklund type relations, i.e. the local equations of motion. Finally, we consider the case where two different theories are separated by a suitable defect matrix. We choose the Liouville theory and the free massless theory and we are able to identify the defect (Darboux) matrix that provides the celebrated hetero-BT for the Liouville theory.

\section{The deformed oscillator}
\label{sec:disc}

Before we consider the model of the deformed oscillator -- with the Lax operator given by \eqref{eq:lax1} -- in the presence of defects, let us first briefly review the model without any discontinuities present. Recall the underlying algebraic structure of the model defined by \eqref{eq:fund1}. Indeed, from this Poisson structure one immediately extracts the following algebraic relations for the discrete fields:
\begin{equation}
\begin{aligned}
	\big\lbrace \aa_n, \aa_m \big\rbrace &= \big\lbrace \ab_n, \ab_m \big\rbrace = \big\lbrace v_n, v_m \big\rbrace = 0, \\
	\big\lbrace \aa_n, v_m \big\rbrace &= \aa_n v_n \delta_{nm}, \\
	\big\lbrace \ab_n, v_m \big\rbrace &= -\ab_n v_n \delta_{nm}, \\
	\big\lbrace \aa_n, \ab_m \big\rbrace &= -2v_n^2 \delta_{nm}.
\end{aligned}
\end{equation}

Building the $ N $ site monodromy matrix:
\begin{equation}
	T(\lambda) = L_{N}(\lambda) L_{N - 1}(\lambda) ... L_{1}(\lambda),
\end{equation}

\noindent the integrals of motion can be read from the expansion about powers of $ \e{\lambda} $ of the generating functional $ \mathcal{G}(\lambda) = \ln{\text{tr } T(\lambda)} $. The first three integrals of motion are then given as:
\begin{equation}
\begin{aligned}
	I^{(0)} &= \sum_{j = 1}^{N} \ln{v_j}, \qquad\quad I^{(1)} = 0, \\
	I^{(2)} &= \sum_{j = 1}^{N} \abv_{j + 1} \aav_{j} - \sum_{j = 1}^{N} v_{j}^{-2},
\end{aligned}
\end{equation}

\noindent where $ \aav_j = \aa_j v_j^{-1} $ and $ \abv_j = \ab_j v_j^{-1} $, for which it immediately follows that:
\begin{equation}
\begin{aligned}
	\big\lbrace \aav_n, \aav_m \big\rbrace &= \big\lbrace \abv_n, \abv_m \big\rbrace = 0, \\
	\big\lbrace \aav_n, v_m \big\rbrace &= \aav_n v_n \delta_{nm}, \\
	\big\lbrace \abv_n, v_m \big\rbrace &= -\abv_n v_n \delta_{nm}, \\
	\big\lbrace \aav_n, \abv_m \big\rbrace &= -(2 + \aav_n \abv_n ) \delta_{nm}.
\end{aligned}
\end{equation}

\noindent It is worth noting that the latter relations coincide with the ones appearing in the Ablowitz-Ladik model \cite{ablow} up to a suitable rescaling of the fields and an appropriate definition of the field $ v_n $ in terms of $ \aav_n,\ \abv_n $. This is an interesting observation that provides a clear connection between the two models.

In addition to the local integrals of motion, one may also derive the time component of the Lax pair using the classical algebra \cite{sts}:
\begin{equation}
	\mathbb{A}_j(\lambda, \mu) = t^{-1}(\lambda) \text{ tr}_a \big\lbrace T_a(N, j, \lambda) r_{ab}(\lambda - \mu) T_a(j - 1, 1, \lambda) \big\rbrace,
\end{equation}

\noindent The $ \mathbb{A}_j^{(m)} $ associated to each of the $ I^{(m)} $, as found by expanding $ \mathbb{A}_j $ about powers of $ \e{\lambda} $, can be seen to be:
\begin{equation}
\begin{aligned}
	\mathbb{A}_j^{(0)} &= \left( \begin{matrix} 1 & 0 \\ 0 & 0 \end{matrix} \right), \qquad\qquad \mathbb{A}_j^{(1)} = 0, \\
	\mathbb{A}_j^{(2)} &= \left( \begin{matrix}
		2\e{2\mu} - \abv_j \aav_{j - 1} & 2\e{\mu} \abv_j \\
		2\e{\mu} \aav_{j - 1} & \abv_j \aav_{j - 1}
	\end{matrix} \right).
\end{aligned} \label{eq:lax2}
\end{equation}

Using the zero-curvature condition on $ \mathbb{A}_j^{(2)} $ (or equivalently, and as a consistency check, Hamilton's equations on $ I^{(2)} $), the equations of motions, i.e. the time evolution of the fields $ \aa_j $, $ \ab_j $, and $ v_j $, can be shown to be:
\begin{equation}
\begin{aligned}
	\dot{\aa}_j &= 2 \aav_{j - 1} v_j - 2 \aav_j v_j^{-1} + \abv_{j + 1} \aav_j \aa_j + \abv_j \aav_{j - 1} \aa_j, \\
	\dot{\ab}_j &= -2 \abv_{j + 1} v_j + 2 \abv_j v_j^{-1} - \abv_{j + 1} \aav_j \ab_j - \abv_j \aav_{j - 1} \ab_j, \\
	\dot{v}_j &= \abv_{j + 1} \aa_j - \ab_j \aav_{j - 1}.
\end{aligned}
\end{equation}

The next subsection is devoted to the study of the discrete system in the presence of local defects.

\subsection{Implementing local defects}

The local defect essentially modifies the monodromy matrix. In particular, the presence of a defect on the $ n $th site of the one dimensional lattice modifies the monodromy matrix as follows:
\begin{equation}
	T(\lambda) = L_{0N}(\lambda) \ldots \tilde{L}_{0n}(\lambda - \theta) \ldots L_{01}(\lambda).
\end{equation}

We choose to consider the following generic $ \tilde{L} $ defect matrix, which satisfies the same classical algebra \eqref{eq:fund1} as $ L $, so that integrability is guaranteed:
\begin{equation}
	\tilde{L}_n = \left( \begin{matrix}
		\e{\lambda - \theta} X_n - \e{-\lambda + \theta} X_n^{-1} & \zb_n \\
		\zz_n & \e{\lambda - \theta} X_n^{-1} - \e{-\lambda + \theta} X_n
	\end{matrix} \right).
\end{equation}

\noindent In fact, this is the key requirement at the algebraic level so that integrability {\it \`{a} la} Liouville is ensured by construction.

As the quadratic algebra is ultralocal, the Poisson bracket of the bulk fields with the defect fields are immediately zero, and the relations between the defect fields are:
\begin{equation}
\begin{aligned}
	\big\lbrace \zz_n, \zz_n \big\rbrace &= \big\lbrace \zb_n, \zb_n \big\rbrace = \big\lbrace X_n, X_n \big\rbrace = 0, \\
	\big\lbrace \zz_n, X_n \big\rbrace &= \zz_n X_n, \\
	\big\lbrace \zb_n, X_n \big\rbrace &= -\zb_n X_n, \\
	\big\lbrace \zz_n, \zb_n \big\rbrace &= 2(X_n^{-2} - X_n^2).
\end{aligned} \nonumber
\end{equation}

The introduction of this defect changes the associated integrals of motion to (with $ \zzX_n = \zz_n X_n^{-1} $ and $ \zbX_n = \zb_n X_n^{-1} $):
\begin{equation}
\begin{aligned}
	\tilde{I}^{(0)} &= \sum_{j \neq n}^{N} \ln{v_j} + \ln{X_n} - \theta, \qquad\qquad \tilde{I}^{(1)} = 0, \\
	\tilde{I}^{(2)} &= \sum_{j \neq n, n - 1}^{N} \abv_{j + 1} \aav_j - \sum_{j \neq n}^{N} v_j^{-2} + \e{\theta} (\zbX_n \aav_{n - 1} + \abv_{n + 1} \zzX_n) + \abv_{n + 1} \aav_{n - 1} X_n^{-2} - \e{2\theta} X_n^{-2}.
\end{aligned} \nonumber
\end{equation}

Expressions for the time component of the Lax pair when defects are present has been derived in \cite{doikou0}. While the defect does leave the 0th and 1st order $ \mathbb{A}_j $ matrices unchanged, as well as $ \mathbb{A}_j^{(2)} $ when $ j \neq n, n + 1 $, around the defect they become slightly ``deformed'' compared to the bulk quantities given in \eqref{eq:lax2}. Introducing a new pair of variables, $ \tilde{\aav}_{n, n - 1} = \e{\theta} \zzX_n + \aav_{n - 1} X_n^{-2} $ and $ \tilde{\abv}_{n, n + 1} = \e{\theta} \zbX_n + \abv_{n + 1} X_n^{-2} $, these are:
\begin{equation}
\begin{gathered}
	\tilde{\mathbb{A}}_n^{(2)} = \left( \begin{matrix}
		2\e{2\mu} - \tilde{\abv}_{n, n + 1} \aav_{n - 1} & 2\e{\mu} \tilde{\abv}_{n, n + 1} \\
		2\e{\mu} \aav_{n - 1} & \tilde{\abv}_{n, n + 1} \aav_{n - 1}
	\end{matrix} \right), \\
	\tilde{\mathbb{A}}_{n + 1}^{(2)} = \left( \begin{matrix}
		2\e{2\mu} - \abv_{n + 1} \tilde{\aav}_{n, n - 1} & 2\e{\mu} \abv_{n + 1} \\
		2\e{\mu} \tilde{\aav}_{n, n - 1} & \abv_{n + 1} \tilde{\aav}_{n, n - 1}
	\end{matrix} \right).
\end{gathered}
\end{equation}

Around the defect point, the equations of motion are also altered to account for the presence of the defect. Hence the local equations of motion are given as:
\begin{equation}
\begin{aligned}
	\dot{\aa}_{n - 1} &= 2 \aav_{n - 2} v_{n - 1} - 2 \aav_{n - 1} v_{n - 1}^{-1} + \tilde{\abv}_{n, n + 1} \aav_{n - 1} \aa_{n - 1} + \abv_{n - 1} \aav_{n - 2} \aa_{n - 1}, \\
	\dot{\ab}_{n - 1} &= -2 \tilde{\abv}_{n, n + 1} v_{n - 1} + 2 \abv_{n - 1} v_{n - 1}^{-1} - \tilde{\abv}_{n, n + 1} \aav_{n - 1} \ab_{n - 1} - \abv_{n - 1} \aav_{n - 2} \ab_{n - 1}, \\
	\dot{v}_{n - 1} &= \tilde{\abv}_{n, n + 1} \aa_{n - 1} - \ab_{n - 1} \aav_{n - 2}, \\
	\dot{\aa}_{n + 1} &= 2 \tilde{\aav}_{n, n - 1} v_{n + 1} - 2 \aav_{n + 1} v_{n + 1}^{-1} + \abv_{n + 2} \aav_{n + 1} \aa_{n + 1} + \abv_{n + 1} \tilde{\aav}_{n, n - 1} \aa_{n + 1}, \\
	\dot{\ab}_{n + 1} &= -2 \abv_{n + 2} v_{n + 1} + 2 \abv_{n + 1} v_{n + 1}^{-1} - \abv_{n + 2} \aav_{n + 1} \ab_{n + 1} - \abv_{n + 1} \tilde{\aav}_{n, n - 1} \ab_{n + 1}, \\
	\dot{v}_{n + 1} &= \abv_{n + 2} \aa_{n + 1} - \ab_{n + 1} \tilde{\aav}_{n, n - 1}.
\end{aligned}
\end{equation}

\noindent At the defect itself, the equations of motion are instead for $ \zz_n $, $ \zb_n $, and $ X_n $:
\begin{equation}
\begin{aligned}
	\dot{\zz}_n &= 2 \e{\theta} \aav_{n - 1} X_n - 2 \e{\theta} \tilde{\aav}_{n, n - 1} X_n^{-1} + \abv_{n + 1} \tilde{\aav}_{n, n - 1} \zz_n + \tilde{\abv}_{n, n + 1} \aav_{n - 1} \zz_n, \\
	\dot{\zb}_n &= -2 \e{\theta} \abv_{n + 1} X_n + 2 \e{\theta} \tilde{\abv}_{n, n + 1} X_n^{-1} - \abv_{n + 1} \tilde{\aav}_{n, n - 1} \zb_n - \tilde{\abv}_{n, n + 1} \aav_{n - 1} \zb_n, \\
	\dot{X}_n &= \e{\theta} \abv_{n + 1} \zz_n - \e{\theta} \zb_n \aav_{n - 1}.
\end{aligned}
\end{equation}

Note that the set of equations becomes increasingly involved as we consider higher integrals of motion. The main difference compared to the continuous case, as will be evident below, is that in discrete models no continuity conditions exist to keep the effect of the discontinuity ``local''. Clearly, the effect of the impurity spreads rabidly when considering higher orders of the hierarchy, providing more and more complicated clusters of equations that describe the time evolution of the defect. We have several examples already at our disposal of similar intricate behaviour, i.e. the Discrete NLS \cite{doikou0}, the Toda chain \cite{doikou-backl}, and the deformed oscillator presented here, which we hope to study under some kind of unified scheme.

\section{The Liouville theory}
\label{sec:cts}

As in the discrete case we shall first review the Liouville model without discontinuities. In fact, one could study the Liouville theory as a certain limit of the sine-Gordon model, however there are certain intricacies (especially in the presence of a defect) that one has to consider when studying the model in more detail. This is also the case when dealing with the associated B\"{a}cklund type relations. For instance in the sine-Gordon case, type I defects give rise to the familiar B\"{a}cklund relations, whereas in the Liouville theory they provide rather trivial results.

Recall the Lax pair for the Liouville model, which may be seen as a suitable massless limit of the sine-Gordon model \cite{FT}:
\begin{equation}
\begin{aligned}
	U(\lambda) &= \frac{1}{2} \left( \begin{matrix}
		-i\pi & -2\e{-\lambda - i\phi} \\
		4\, \text{sinh}(\lambda - i\phi) & i\pi
	\end{matrix} \right), \\
	V(\lambda) &= \frac{1}{2} \left( \begin{matrix}
		-i\phi_x & 2\e{-\lambda - i\phi} \\
		4\, \text{cosh}(\lambda - i\phi) & i\phi_x
	\end{matrix} \right),
\end{aligned} \label{eq:LaxLiouv}
\end{equation}

\noindent where the fields are $ \phi(x, t) $ and $ \pi(x, t) = \phi_t $. Inserting these into the zero curvature condition ($ U_t - V_x + [U, V] = 0 $) returns the Liouville equation:
\begin{equation}
	\phi_{tt} - \phi_{xx} - 4i\e{-2i\phi} = 0.
\end{equation}

After performing a gauge transformation $ T \to g^{-1} T g $, where $ g = \e{-i \phi \sigma^z / 2} $ and $ T $ is the monodromy matrix built below, the space component of the Lax pair looks like:
\begin{equation}
	\tilde U(\lambda) = \frac{1}{2} \left( \begin{matrix}
		i(\phi_x - \pi) & -2\e{-\lambda} \\
		2\e{\lambda - 2i \phi} - 2\e{-\lambda} & -i(\phi_x - \pi)
	\end{matrix} \right).
\end{equation}

Using this, an infinite series of integrals of motion can be found from the expansion of the generating functional $ \mathcal{G} = \ln{\text{tr } T(L, -L, \lambda)} $ about its powers of $ \e{\lambda} $, where $ T $ is the monodromy matrix given by:
\begin{equation}
	\tilde T(x, y, \lambda) = \Pexp{\int_{y}^{x} \tilde U(\xi, \lambda) \dd \xi},
\end{equation}

\noindent which can also be split as:
\begin{equation}
	\tilde T(x, y, \lambda) = \big( \mathbb{I} + W(x, \lambda) \big) \e{Z(x, y, \lambda)} \big( \mathbb{I} + W(y, \lambda) \big)^{-1},
\end{equation}

\noindent where $ Z(x, y, \lambda) $ and $ W(x, \lambda) $ are diagonal and anti-diagonal matrices respectively, that can be expanded in powers of $ u = \e{\lambda} $:
\begin{equation}
\begin{aligned}
	W(x, \lambda) &= \sum_{n = 1}^{\infty} u^n W^{(n)}(x), \\
	Z(x, y, \lambda) &= \sum_{n = -1}^{\infty} u^n Z^{(n)}(x, y).
\end{aligned}
\end{equation}

Even after this splitting, $ T $ must still obey the auxiliary linear problem, so by inserting this definition of $ T $ into the spatial part (that $ T_x = U T $) and separating the resulting equation into its diagonal and anti-diagonal parts we can get expressions for the $ Z^{(n)} $ and $ W^{(n)} $:
\begin{equation}
\begin{aligned}
	&\sum_{n = -1}^{\infty} u^n Z^{(n)}_x = U_D + \sum_{n = 0}^{\infty} u^n U_A W^{(n)}, \\
	&\sum_{n = 0}^{\infty} u^n W^{(n)}_x + \sum_{n = 0}^{\infty} u^n \big[ W^{(n)}, U_D \big] - U_A + \sum_{n, m = 0}^{\infty} u^{n + m} W^{(n)} U_A W^{(m)} = 0,
\end{aligned} \label{eqs:WandZ}
\end{equation}

\noindent where $ U_D $ and $ U_A $ are the diagonal and anti-diagonal parts of $ U $ respectively.

As the trace is cyclic, when $ \mathcal{G} $ is written in terms of the $ W $ and $ Z $ matrices the $ (\mathbb{I} + W(L)) $ and $ (\mathbb{I} + W(-L))^{-1} $ terms cancel out (after imposing periodic boundary conditions), leaving:
\begin{equation}
	\mathcal{G} = \ln{\text{tr } \e{Z(L, -L, \lambda)}}.
\end{equation}

\noindent The matrix exponential of a diagonal matrix is just the matrix with its elements exponentiated, and in the limit as $ u^{-1} \to \infty $, the dominant term will be $ Z_{11}^{(-1)} $, so the term depending on $ Z_{22} $ can be dropped, leaving:
\begin{equation}
	\mathcal{G} = u^{-1} Z^{(-1)}_{11} + Z^{(0)}_{11} + u Z^{(1)}_{11} + ....
\end{equation}

The first (non-trivial) integral of motion is the one of order $ u $, so is just the top-left element of $ Z^{(1)} $, i.e.:
\begin{equation}
	I^{(1)} = -\frac{1}{2} \int_{-L}^{L} \bigg( \frac{1}{4} \big( \phi_x^2 + \pi^2 - 2 \phi_x \pi \big) + \e{-2i \phi} \bigg) \dd x.
\end{equation}

To find the Hamiltonian for the system, a second integral of motion needs to be found by exploiting the symmetry that $ U(\phi, \pi, \lambda) = \sigma^x U^T(\phi, -\pi, \lambda) \sigma^x $. Then, as $ \sigma^x \sigma^x = \mathbb{I} $, the $ \sigma^x $ matrices can be pulled out of the exponential, leaving the monodromy matrix found after the symmetry as:
\begin{equation}
	T(x, y, \phi, \pi, \lambda) = \sigma^x \left( \Pexp{\int_y^x U^T(\xi, \phi, -\pi, \lambda) \dd \xi} \right) \sigma^x.
\end{equation}

\noindent The effect this has is that after the $ W^{(n)} $ and $ Z^{(n)} $ matrices are modified accordingly, another integral of motion can be read from the top-left entry of $ Z^{(1)} $:
\begin{equation}
	I^{(1)}_{\text{sym}} = -\frac{1}{2} \int_{-L}^{L} \bigg( \frac{1}{4} \big( \phi_x^2 + \pi^2 + 2 \phi_x \pi \big) + \e{-2i \phi} \bigg) \dd x.
\end{equation}

\noindent As any combination of integrals of motion must itself be an integral of motion, the momentum and Hamiltonian of this system can be found from these by taking their difference and sum, respectively:
\begin{equation}
\begin{aligned}
	I^{(1)}_{\text{sym}} - I^{(1)} \propto \mathcal{P} &= \int_{-L}^{L} \phi_x \pi \dd x, \\
	I^{(1)}_{\text{sym}} + I^{(1)} \propto \mathcal{H} &= \int_{-L}^{L} \bigg( \frac{1}{2} \big( \phi_x^2 + \pi^2 \big) + 2e^{-2i \phi} \bigg) \dd x.
\end{aligned}
\end{equation}

\-

As discussed in \cite{ACDK} one can consider the ``dual'' picture, i.e. study the system along the time direction. Then the monodromy matrix is made to satisfy the time half of the auxiliary linear problem instead of the space half, so becomes the exponential of the time component of the Lax pair $ V $. Repeating the calculations with $ T(x, y, \lambda) = \Pexp{\int_{y}^{x} V \dd t} $ gives the integral of motion as:
\begin{equation}
	I^{(1)} = -\frac{1}{2} \int_{-L}^{L} \bigg( \frac{1}{4} \big( \phi_x^2 + \pi^2 - 2 \phi_x \pi \big) - \e{-2i \phi} \bigg) \dd t.
\end{equation}

\noindent This time, the symmetry that $ V(\phi, \phi_x, \lambda) = \sigma^x V^T(\phi, -\phi_x, \lambda) \sigma^x $ is used, which gives the alternate integral of motion as:
\begin{equation}
	I^{(1)}_{\text{sym}} = -\frac{1}{2} \int_{-L}^{L} \bigg( \frac{1}{4} \big( \phi_x^2 + \pi^2 + 2 \phi_x \pi \big) - \e{-2i \phi} \bigg) \dd t,
\end{equation}

\noindent so the ``time versions'' of the momentum and Hamiltonian are:
\begin{equation}
\begin{aligned}
	I^{(1)}_{\text{sym}} - I^{(1)} \propto \mathcal{P}^{(t)} &= \int_{-L}^{L} \phi_x \pi \dd x, \\
	I^{(1)}_{\text{sym}} + I^{(1)} \propto \mathcal{H}^{(t)} &= \int_{-L}^{L} \bigg( \frac{1}{2} \big( \phi_x^2 + \pi^2 \big) - 2\e{-2i \phi} \bigg) \dd t.
\end{aligned}
\end{equation}

In fact, the duality is in general rather straightforward for the Liouville, sine-Gordon, and Affine Toda Field theory cases. This is due to the fact that all the mentioned models are relativistic, thus an easy ``dictionary'' that leads from the space-like to the time-like picture can be used:
\begin{equation}
	x \to t, ~~~~\phi_t \to \phi_x, ~~~~~\lambda \to \lambda+ \frac{i \pi}{2}. \label{eq:dict}
\end{equation}

\subsection{Implementing defects}

As in the case with the deformed oscillator, a type \rm{II} defect is considered:
\begin{equation}
	\tilde{L}(\lambda) = \left( \begin{matrix}
		\e{\lambda} X - \e{-\lambda} X^{-1} & \zb \\
		\zz & \e{\lambda} X^{-1} - \e{-\lambda} X
	\end{matrix} \right).
\end{equation}

This defect is inserted at a point $ x_0 \in [-L, L] $, flanked by the points $ x_{0}^{+} > x_0 > x_{0}^{-} $, such that the monodromy matrix now takes the form:
\begin{equation}
	T(L, -L, \lambda) = T^{+}(L, x_{0}^{+}, \lambda) \tilde{L}(x_0, \lambda) T^{-}(x_{0}^{-}, -L, \lambda),
\end{equation}

\noindent with $ T^{\pm} $ satisfying their own auxiliary linear problems, $ T^{\pm}_x = U^{\pm} T^{\pm} $ with their own Lax pair matrices, given by (after performing the gauge transformation given above):
\begin{equation}
	\tilde U^{\pm}(\lambda) = \frac{1}{2} \left( \begin{matrix}
		i(\phi^{\pm}_x - \pi^{\pm}) & -2\e{-\lambda} \\
		2\e{\lambda - 2i \phi^{\pm}} - 2\e{-\lambda} & -i(\phi^{\pm}_x - \pi^{\pm})
	\end{matrix} \right).
\end{equation}

\noindent Each of these $ T^{\pm} $ can be split into a combination of $ W^{\pm (n)} $ and $ Z^{\pm (n)} $, which satisfy the relations \eqref{eqs:WandZ}. The $ W^{\pm (n)} $ and $ Z^{\pm (n)} $ matrices will therefore take similar forms to those without the defect.

Again, the integrals of motion are read from the expansion of $ \mathcal{G} = \ln{\text{tr } T(L, -L, \lambda) } $ about powers of $ u = \e{\lambda} $, so assuming periodic boundary conditions, this is (suppressing the $ \lambda $ dependence):
\begin{equation}
	\mathcal{G} = \ln \bigg[ \text{tr } \Big\lbrace \e{Z^{+}(L, x_{0}^{+})} \big( \mathbb{I} + W^{+}(x_{0}^{+}) \big)^{-1} \big( g^{+} (x_{0}^{+}) \big)^{-1} \tilde{L}\, g^{-} (x_{0}^{-})  \big( \mathbb{I} + W^{-}(x_{0}^{-}) \big) \e{Z^{-}(x_{0}^{-}, -L)} \Big\rbrace \bigg], \nonumber
\end{equation}

\noindent where $ g^{\pm} $ are the gauge transformations for $ T^{\pm} $ defined in the case without defects. Considering again the limit as $ u^{-1} \to \infty $, the elements $ Z^{\pm}_{11} $ will dominate over $ Z^{\pm}_{22} $, so the exponentials of the $ Z^{\pm} $ matrices can be replaced:
\begin{equation}
	\e{Z^{\pm}} \to \e{Z^{\pm}_{11}} \left( \begin{matrix} 1 & 0 \\ 0 & 0 \end{matrix} \right), \nonumber
\end{equation}

\noindent then as the trace of these matrices is considered, the generating functional can be written (after factoring out the exponentials of $ Z^{\pm}_{11} $):
\begin{equation}
	\mathcal{G} = Z^{+}_{11} + Z^{-}_{11} + \ln  \bigg[ \big( \mathbb{I} + W^{+}(x_{0}^{+}) \big)^{-1} \big( g^{+} (x_{0}^{+}) \big)^{-1} \tilde{L}(x_0)  g^{-} (x_{0}^{-})  \big( \mathbb{I} + W^{-}(x_{0}^{-}) \big) \bigg]_{11}. \nonumber
\end{equation}

Explicitly calculating the term inside the logarithm, the coefficient of $ u^1 $ is the first (non-trivial) integral of motion:
\begin{align}
	I^{(1)} &= -\frac{1}{2} \int_{x_{0}^{+}}^{L} \Big( \frac{1}{4} \big( \phi^{+}_x - \pi^{+} \big)^2 + \e{-2i \phi^{+}} \Big) \dd x - \frac{1}{2} \int_{-L}^{x_{0}^{-}} \Big( \frac{1}{4} \big( \phi^{-}_x - \pi^{-} \big)^2 + \e{-2i \phi^{-}} \Big) \dd x \nonumber \\
	&\qquad+ \frac{1}{\mathcal{D}} \Big( \zz \e{-\frac{i}{2} (\phi^{+} + \phi^{-})} + \zb \e{\frac{i}{2} (\phi^{+} + \phi^{-})} \Big) - \frac{i\mathcal{A}}{2\mathcal{D}} \Big( \phi^{+}_x - \pi^{+} + \phi^{-}_x - \pi^{-} \Big) + \frac{i}{2} \Big( \phi^{+}_x - \pi^{+} \Big), \nonumber\\ &
\end{align}

\noindent where $ \mathcal{D} $ and $ \mathcal{A} $ are defined to be:
\begin{equation}
\begin{aligned}
	\mathcal{D} &= X \e{-\frac{i}{2} (\phi^{+} - \phi^{-})} + X^{-1} \e{\frac{i}{2} (\phi^{+} - \phi^{-})}, \\
	\mathcal{A} &= X \e{-\frac{i}{2} (\phi^{+} - \phi^{-})}.
\end{aligned}
\end{equation}

In the presence of the defect $ \tilde{L} $, the symmetry of the monodromy matrix can instead be written as (suppressing all unchanged dependencies):
\begin{equation}
	T \Big( U(\pi), \tilde{L}(X) \Big) = T \Big( \sigma^x U^T(-\pi) \sigma^x, \sigma^x \tilde{L}^T(X^{-1}) \sigma^x \Big).
\end{equation}

\noindent As the symmetry affects the untransformed Lax matrix, the gauge transformation is not performed until after the monodromy matrices are changed under the symmetry, so can be written in terms of the $ W^{\pm}_{\text{sym}} $ and $ Z^{\pm}_{\text{sym}} $ from the case above as:
\begin{equation}
\begin{aligned}
	\tilde T^{+}_{\text{sym}} &= \big( g^{+} (L) \big) \sigma^x \big( \mathbb{I} + W^{+}_{\text{sym}}(L) \big) \e{Z^{+}_{\text{sym}}(L, x_{0}^{+})} \big( \mathbb{I} + W^{+}_{\text{sym}}(x_{0}^{+}) \big)^{-1} \sigma^x \big( g^{+} (x_{0}^{+}) \big)^{-1},\\
	\tilde T^{-}_{\text{sym}} &= \big( g^{-} (x_{0}^{-}) \big) \sigma^x \big( \mathbb{I} + W^{-}_{\text{sym}}(x_{0}^{-}) \big) \e{Z^{-}_{\text{sym}}(x_{0}^{-}, -L)} \big( \mathbb{I} + W^{-}_{\text{sym}}(-L) \big)^{-1} \sigma^x \big( g^{-} (-L) \big)^{-1}.
\end{aligned} \nonumber
\end{equation}

\noindent Therefore, assuming periodic boundary conditions and noticing that $ \sigma^x g^{\pm} \sigma^x = (g^{\pm})^{-1} $, the generating functional can be modified accordingly. In the $ u^{-1} \to \infty $ limit of this $ \mathcal{G} $, the first non-trivial integral of motion arises from the $ u^1 $ term as:
\begin{equation}
\begin{aligned}
	I^{(1)}_{\text{sym}} &= -\frac{1}{2} \int_{x_{0}^{+}}^{L} \Big( \frac{1}{4} \big( \phi^{+}_x + \pi^{+} \big)^2 + \e{-2i \phi^{+}} \Big) \dd x - \frac{1}{2} \int_{-L}^{x_{0}^{-}} \Big( \frac{1}{4} \big( \phi^{-}_x + \pi^{-} \big)^2 + \e{-2i \phi^{-}} \Big) \dd x \\
	&\quad+ \frac{1}{\mathcal{D}} \Big( \zz \e{-\frac{i}{2} (\phi^{+} + \phi^{-})} + \zb \e{\frac{i}{2} (\phi^{+} + \phi^{-})} \Big) - \frac{i\mathcal{A}^{-1}}{2\mathcal{D}} \Big( \phi^{+}_x + \pi^{+} + \phi^{-}_x + \pi^{-} \Big) + \frac{i}{2} \Big( \phi^{+}_x + \pi^{+} \Big).
\end{aligned}
\end{equation}

Combining this with the pre-symmetry integral of motion in the same manner as was done without the defect gives the momentum (proportional to $ I^{(1)}_{\text{sym}} - I^{(1)} $) and the Hamiltonian (proportional to $ I^{(1)}_{\text{sym}} + I^{(1)} $) in the presence of the defect as:
\begin{align}
	\mathcal{P} &= \int_{x_{0}^{+}}^{L} \phi^{+}_x \pi^{+} \dd x + \int_{-L}^{x_{0}^{-}} \phi^{-}_x \pi^{-} \dd x - i \big( \pi^{+} - \pi^{-} \big) - i \frac{A - A^{-1}}{\mathcal{D}} \big( \phi^{+}_x + \phi^{-}_x \big), \nonumber \\
	\mathcal{H} &= \int_{x_{0}^{+}}^{L} \Big( \frac{1}{2} \big( (\phi^{+}_x)^2 + (\pi^{+})^2 \big) + 2\e{-2i \phi^{+}} \Big) \dd x + \int_{-L}^{x_{0}^{-}} \Big( \frac{1}{2} \big( (\phi^{-}_x)^2 + (\pi^{-})^2 \big) + 2\e{-2i \phi^{-}} \Big) \dd x \nonumber \\
	&\qquad- \frac{4}{\mathcal{D}} \Big( \zz \e{-\frac{i}{2} (\phi^{+} + \phi^{-})} + \zb \e{\frac{i}{2} (\phi^{+} + \phi^{-})} \Big) - i \big( \phi^{+}_x - \phi^{-}_x \big) - i \frac{A - \mathcal{A}^{-1}}{\mathcal{D}} \big( \pi^{+} + \pi^{-} \big).
\nonumber
\end{align}

\noindent Note that, as in the bulk case, we can establish the ``dual'' time-like picture in a straightforward manner based on the dictionary \eqref{eq:dict} and considering the same type of defect along the time direction.

\subsubsection*{Lax pairs $\&$ gluing conditions}

The derivation of the time components of the Lax pairs around the defect point is a necessary step according to the analysis in \cite{avan-doikou} for establishing the related gluing conditions around the defect point. It turns out that these relations are also automatically satisfied when extracting the B\"{a}cklund transformation relations.

The expression for the time components of the Lax pair around the defect are found from the left and right via the equations derived in \cite{avan-doikou}:
\begin{equation}
\begin{aligned}
	\tilde{V}^{+}(x_0, \lambda, \mu) &= t^{-1}(\lambda) \text{tr}_a \left\lbrace T_{a}^{+}(L, x_0, \lambda) r_{ab}(\lambda - \mu) \tilde{L}_{a}(x_0, \lambda) T_{a}^{-}(x_0, -L, \lambda) \right\rbrace, \\
	\tilde{V}^{-}(x_0, \lambda, \mu) &= t^{-1}(\lambda) \text{tr}_a \left\lbrace T_{a}^{+}(L, x_0, \lambda) \tilde{L}_{a}(x_0, \lambda) r_{ab}(\lambda - \mu) T_{a}^{-}(x_0, -L, \lambda) \right\rbrace.
\end{aligned}
\end{equation}

The bulk quantities provide the familiar left and right Liouville theory Lax pairs, and these two bulk $ V $ matrices with $ r_{ab} $ away from the defect are required to smoothly transition to the above matrices with tildes (those with $ r_{ab} $ at the defect) in the limit as $ x \to x_{0}^{\pm} $. This requirement will give rise to certain sewing conditions.

After splitting these matrices about powers of $ u $, the first order terms away from the defect are:
\begin{equation}
\begin{aligned}
	V^{+(1)}(x, \mu) &= -i \sigma^z \big( \phi^{+}_{x}(x) - \pi^{+}(x) \big) + 4 \e{-\mu} \big( \sigma^{+} \e{-i\phi^{+}} + \sigma^{-} \e{i\phi^{+}} \big), \\
	V^{-(1)}(x, \mu) &= -i \sigma^z \big( \phi^{-}_{x}(x) - \pi^{-}(x) \big) + 4 \e{-\mu} \big( \sigma^{+} \e{-i\phi^{-}} + \sigma^{-} \e{i\phi^{-}} \big),
\end{aligned}
\end{equation}

\noindent and those near the defect are:
\begin{equation}
\begin{aligned}
	\tilde{V}^{+(1)}(x_0, \mu) &= \frac{1}{\mathcal{D}^2} \sigma^{z} \Big( \mathcal{A}^{-1} \zz \e{-\frac{i}{2}(\phi^{+} + \phi^{-})} - \mathcal{A} \zb \e{\frac{i}{2}(\phi^{+} + \phi^{-})} - \frac{i}{2} \big( \phi^{+}_{x} - \pi^{+} + \phi^{-}_{x} - \pi^{-} \big) \Big) \\
	&\qquad+ \frac{2}{\mathcal{D}} \e{-\mu} \big( \sigma^{+} X^{-1} \e{-\frac{i}{2} (\phi^{+} + \phi^{-})} + \sigma^{-} X \e{\frac{i}{2} (\phi^{+} + \phi^{-})} \big), \\
	\tilde{V}^{-(1)}(x_0, \mu) &= \frac{1}{\mathcal{D}^2} \sigma^{z} \Big( \mathcal{A}^{-1} \zb \e{\frac{i}{2}(\phi^{+} + \phi^{-})} - \mathcal{A} \zz \e{-\frac{i}{2}(\phi^{+} + \phi^{-})} - \frac{i}{2} \big( \phi^{+}_{x} - \pi^{+} + \phi^{-}_{x} - \pi^{-} \big) \Big) \\
	&\qquad+ \frac{2}{\mathcal{D}} \e{-\mu} \big( \sigma^{+} X \e{-\frac{i}{2} (\phi^{+} + \phi^{-})} + \sigma^{-} X^{-1} \e{\frac{i}{2} (\phi^{+} + \phi^{-})} \big).
\end{aligned}
\end{equation}

From requiring that $ V^{\pm (1)} \to \tilde{V}^{\pm (1)} $ as $ x \to x_{0} $, the anti-diagonal elements give the first sewing condition $ S_1 $:
\begin{equation}
	S_1 = X - \e{\frac{i}{2}(\phi^{+} - \phi^{-})},
\end{equation}

\noindent such that if the condition is satisfied (i.e. if $ X = \e{\frac{i}{2}(\phi^{+} - \phi^{-})} $), then $ S_1 \approx 0 $.

\subsubsection*{B\"acklund type relations}

We shall derive below the time and space B\"{a}cklund type relations (we refer the interested reader to \cite{wahlq}--\cite{sklyanin-back} for more details on B\"{a}cklund transformations) arising from the local equations of motion on the defect point. First, the case with space-like defects is considered and the $ t $-part of the B\"{a}cklund transformation is obtained. Then, by similar reasoning, the study of the integrable defect along the time axis will provide the $ x $-part of the B\"{a}cklund transformation.

Using a type \rm{II} Darboux matrix:
\begin{equation}
	L = \left( \begin{matrix}
		u \e{-\theta} X - u^{-1} \e{\theta} X^{-1} & Y \\
		Z & u \e{-\theta} X^{-1} - u^{-1} \e{\theta} X
	\end{matrix} \right), \label{eq:DarbouxMat}
\end{equation}

\noindent and inserting this into the B\"{a}cklund transformation relations:
\begin{equation}
	L_x = \tilde{U} L - L U, \qquad\qquad\quad L_t = \tilde{V} L - L V,
\end{equation}

\noindent where the subscripts represent differentiation with respect to the named variable, gives expressions for each of the fields $ X $, $ Y $, and $ Z $ in terms of the fields $ \phi $, $ \tilde{\phi} $, $ \pi $, and $ \tilde{\pi} $. First, the $ X $ expression is simply:
\begin{equation}
	X = \e{\frac{i}{2}(\tilde{\phi} - \phi)}.
\end{equation}

\noindent This is then used in the $ t $-part of the space-like B\"{a}cklund transformation relations to obtain the following set of equations between $ Y $ and $ Z $. From the diagonal elements:
\begin{equation}
\begin{aligned}
	i(\tilde \phi_t - \phi_t) &= -2Y(\e{\theta} \e{-\frac{i(\phi + \tilde \phi)}{2}} + \e{-\theta} \e{\frac{i(\phi + \tilde \phi)}{2}}) + 2Z\e{-\theta} \e{-\frac{i(\phi + \tilde \phi)}{2}}, \\
	i(\tilde \phi_x - \phi_x) &= -2Y(\e{\theta} \e{-\frac{i(\phi + \tilde \phi)}{2}} - \e{-\theta} \e{\frac{i(\phi + \tilde \phi)}{2}}) - 2Z\e{-\theta} \e{-\frac{i(\phi + \tilde \phi)}{2}},
\end{aligned}
\end{equation}

\noindent and the anti-diagonal elements:
\begin{equation}
\begin{aligned}
	Y_t &= -i(\phi_x + \tilde \phi_x) Y - \e{-\theta} \e{-\frac{i(\phi + \tilde \phi)}{2}} \sinh i(\tilde \phi - \phi), \\
	Z_t &= i(\phi_x + \tilde \phi_x) Z + \e{\theta} \e{-\frac{i(\phi + \tilde \phi)}{2}} \sinh i(\tilde \phi - \phi) + \e{-\theta} \e{\frac{i(\phi + \tilde \phi)}{2}} \sinh i(\tilde \phi - \phi).
\end{aligned}
\end{equation}

\noindent It is worth noting that the latter relations arise also as gluing/analyticity conditions when requiring analyticity of the $ V $ operator around the defect point $ \tilde V(x_0) \to V^{\pm}(x_0) $ as previously discussed. This coincidence is a strong indication of the validity of the process followed to describe the integrable defect.

Similarly, when the time-like defect is considered the anti-diagonal elements of the $ x $-part read as:
\begin{equation}
\begin{aligned}
	Y_x &= -i(\phi_t + \tilde \phi_t) Y + \e{-\theta} \e{-\frac{i(\phi + \tilde \phi)}{2}} \sinh i(\tilde \phi - \phi), \\
	Z_x &= i(\phi_t + \tilde \phi_t) Z + \e{\theta} \e{-\frac{i(\phi + \tilde \phi)}{2}} \sinh i(\tilde \phi - \phi) - \e{-\theta} \e{\frac{i(\phi + \tilde \phi)}{2}} \sinh i(\tilde \phi - \phi).
\end{aligned}
\end{equation}

To describe the effect of the discontinuity on the one dimensional system, one has to, in principal, consider these equations separately. In the space like description, we are basically interested in the time evolution of the defect degrees of freedom encoded in the defect matrix $ L $. The main difference to the familiar B\"{a}cklund transformation relation is that the $ x $ and $ t $ parts are simultaneously satisfied, whereas in the defect picture, depending on whether we consider the space-like or time-like description, we focus on the $ t $ or $ x $ evolution respectively.

\section{Defects as interfaces between models}

Much as the presence of an integrable defect can be thought of as a localised B\"acklund transformation, we can build what would be the equivalent of a hetero-B\"acklund transformation by using an integrable defect to interface between two different models. At the level of the monodromy matrix, when considering two distinct models on the line separated my the defect matrix $ \tilde{L} $ then:
\begin{equation}
	T(\lambda) = T^{+}(\lambda)\ \tilde{L}(x_0, \lambda)\ T^{-}(\lambda). \nonumber
\end{equation}

\noindent but now $ T^{+} $ and $ T^{-} $ are different theories. For integrability to be ensured the following condition should hold (see \cite{doikou-backl} for more details):
\begin{equation}
	\frac{\text{d} \tilde  L(\lambda)}{\text{d} t} = V^{+}\ \tilde L(\lambda) - \tilde L(\lambda )\ V^{-}(\lambda). \label{eq:main1}
\end{equation}

For instance, consider the interface between the free massless theory and the Liouville theory, which is the simplest possible example of an interface between the sine-Gordon model and the Liouville theory. Let us consider a slightly modified Lax pair for the Liouvile theory:
\begin{equation}
	U^+(\lambda) = \frac{1}{2} \left( \begin{matrix}
		-i\tilde \pi & -2c \e{-\lambda + i\tilde \phi} \\
		-2c\, \e{\lambda + i\tilde \phi} & i\pi
	\end{matrix} \right), ~~~~~~~
	V^+(\lambda) = \frac{1}{2} \left( \begin{matrix}
		-i\tilde \phi_x & 2c\e{-\lambda + i\tilde \phi} \\
		2c\, \e{\lambda + i\phi} & i\tilde \phi_x
	\end{matrix} \right),
\label{eq:LaxLiouv2}
\end{equation}

\noindent and the equations of motion from the zero curvature condition read as:
\begin{equation}
	\partial_x^2 \tilde \phi - \partial_t^2 \tilde \phi+ 4ic^2 \e{2i\tilde \phi} = 0. \label{eq:em1}
\end{equation}

\noindent The Lax pair for the free theory is very simple:
\begin{equation}
	U^{-}(\lambda) = -\frac{i}{2} \pi\ {\mathbb I}, ~~~~~~~
	V^{-}(\lambda) = -\frac{i}{2} \phi_x\ {\mathbb I},
\label{eq:Laxfree}
\end{equation}

\noindent where $ {\mathbb I} $ is the $ 2 \times 2 $ unit matrix, and the corresponding equations of motion are:
\begin{equation}
	\partial_x^2 \phi - \partial_t^2 \phi =0. \label{eq:em2}
\end{equation}

\noindent It is clear the the equations of motion remain invariant if we multiply $U^{-}$ and $V^{-}$ with the same constant matrix.

We chose to consider the following Darboux matrix:
\begin{equation}
	\tilde L(\lambda,  \Theta )=  \left( \begin{matrix}
		A & X \e{-\lambda-\Theta}\\
		Z \e{\lambda+\Theta} & B
	\end{matrix} \right),
\end{equation}

\noindent where $\Theta$ is an extra free parameter (B\"acklund transformation parameter) and the elements $A,\  B,\  X,\  Z$ are to be determined via \eqref{eq:main1}. Indeed, setting:
\begin{equation}
	A = X = \e{\frac{i}{2} (\tilde \phi +\phi)}, ~~~~Z = B = \e{\frac{1}{2} (\tilde \phi -\phi)},
\end{equation}

\noindent and using light cone coordinates $z = x + t,\ \bar{z} = x - t$ for convenience, then by solving \eqref{eq:main1} we obtain:
\begin{equation}
\begin{gathered}
	i\partial_z(\tilde \phi -\phi) = -2c \e{\Theta} \e{i (\tilde \phi+\phi)}, \\
	i\partial_{\bar z}(\tilde \phi +\phi) =-2c \e{-\Theta} \e{i (\tilde \phi-\phi)},
\end{gathered} \label{eq:lbt}
\end{equation}

\noindent which is the celebrated hetero-B\"acklund transformation for the Liouville theory, and the solution of the Liouville equation is expressed in terms of the free field. It is also clear via \eqref{eq:lbt} that the fields satisfy the correct equations of motion \eqref{eq:em1}, \eqref{eq:em2}. As expected, the same relations are obtained if one considers the $x$ part of the B\"akclund relations i.e. a time like defect.

\section{Discussion}

Some generic comments can be made for the considered models. The typical property of the discrete and continuum Liouville model is the fact that the $ L $ matrix is somehow ``singular'' and is associated to the harmonic oscillator algebra, as opposed to the case of the generalised XXX and XXZ models and their continuous counterparts, for example.

Another key issue one may address is the classical scattering of solitonic excitations. This can be studied by acting on the auxiliary function with various B\"{a}cklund transformations, and then considering the asymptotic behaviour of the auxiliary function. Effectively, this can be better interpreted by the derivation of the related Gelfand-Levitan-Marchenko equation (see e.g. \cite{FT}) in the presence of local discontinuities via the Zakharov-Shabat dressing (see e.g. \cite{matveev}). Let us stress that particular emphasis should be given when considering the infinite product of fundamental Darboux matrices acting on the defect point.

Although we have been able to identify the Darboux-B\"acklund transformation for the Liouville theory we would like to generalize these computations in order to identify the hetero B\"acklund transformation for the Liouville and sine-Gordon theories as well as consider the discrete analogues of these models.  It is clear that both the continuous and discrete Liouville theories merit further investigation, as wel as their higher rank generalisations. According to the Hamiltonian description \cite{avan-doikou}, the main algebraic requirement is that the bulk theories, as well as the defect matrix, share the same classical algebraic content, i.e. the same $ r $-matrix. These kinds of discontinuities may be seen as some kind of local ``gauge" transformations relating solutions of different nonlinear PDEs.

In general, the derivation of a generic auto- or hetero-B\"{a}cklund transformation is associated to the identification of the general Darboux matrix, formally expressed as:
\begin{equation}
	{\mathbb M}(\lambda) = \sum_{n = -N}^{N} \e{n \lambda} {\mathbb P}_n. \nonumber
\end{equation}
This general Darboux matrix should satisfy the classical fundamental algebra \eqref{eq:fund1}, so by imposing this requirement, the structure of the generic Darboux (and/or defect) matrix can be established. Moreover, this expansion should formally provide the connection with the non-local charges of the theories under study, which essentially encode the underlying symmetry of the model. Consequently, a connection between the generic Darboux transformation and the associated deformed algebra is a really significant question.

\section*{Acknowledgments}

I.F. would like to thank the EPSRC funding council for a PhD studentship.

\end{document}